\documentclass[reprint,fleqn,superscriptaddress,twocolumn,showpacs,
amsmath,amssymb,aps,prb,short bibliography]{revtex4-2}
\usepackage[english]{babel}
\usepackage[all]{xy}
\usepackage{gensymb}
\usepackage{graphicx,psfrag,times,epsfig,color}
\usepackage{verbatim,natbib}
\setcitestyle{super}
\usepackage{booktabs}
\usepackage{color}
\usepackage{makeidx}
\usepackage{amsmath}
\usepackage{bm}
\usepackage{amsfonts}
\usepackage{amssymb}
\usepackage{hyperref}
\usepackage{bm}
\usepackage{ragged2e}

\begin{document}

\title{A unified model of the Hall Effect from insulator to overdoped compounds in cuprate superconductors}
\author{J\'ulia C. Anjos, H\'ercules S. Santana and E. V. L. de Mello\thanks{\email{evlmello@id.uff.br}}
}
\affiliation{Instituto de F\'{\i}sica, Universidade Federal Fluminense, 24210-346 Niter\'oi, RJ, Brazil}


\begin{abstract}

	Measurements of the Hall coefficient in La$_{2-x}$Sr$_x$CuO$_4$, ranging from the undoped ($x = p = 0$) Mott insulator to overdoped compounds,
	exhibit a temperature dependence that offers insights into their electronic structure. We interpret these results using a model based on the
	theory of phase-separation (PS) dynamics, which begins at half-filled ($n = 1$) and at a temperature $T_{\rm PS}(p)$, near the pseudogap temperature
	$T^*(p)$. The $n = 1$ holes have low mobility and provide the modulations of the charge density waves (CDW). As doping
	increases from $p = 0$, these modulations guide the additional p holes to occupy alternating CDW domains.
	This charge inhomogeneity may
	facilitate the formation of localized superconducting amplitudes below the critical onset temperature $T_{\rm c}^{\rm max}(p)$. Using
	thermal activation expressions, along with quantum tunnelling between the charge domains, we successfully reproduce all Hall coefficient
	measurements $R_{\rm H}(p,T)$ and highlight the relevant energies of cuprates.  The calculations confirm three significant
	electronic features: the phase-separating role of the pseudogap temperature, the superconducting state achieved through phase coherence,
	and the two types of charge carriers whose energies and mobilities become comparable at $p \approx 0.19$, where
	$T^*(p) \approx T_{\rm c}^{\rm max}(p)$. This results in a crossover from $n = p$ to $n = 1 + p$. These findings, along with the
	$R_{\rm H}(p,T)$ calculations from insulating to overdoped compounds, underscore the critical role of the electronic
	phase separation in the properties of cuprates.

\end{abstract}
\maketitle
 
\section{Introduction}

High-critical temperature cuprate superconductors (HTS) are derived from a parent compound that is a Mott insulator with one hole per copper (Cu) site ($n = 1$). This localization occurs due to strong on-site Coulomb repulsion, which allows for mobility as doping increases. It was soon established that, aside from exhibiting zero resistivity below the superconducting critical temperature ($T_{\rm{c}}$), HTS show minimal similarities to conventional metallic superconductors. For example, the energy gap that characterizes the superconducting state often does not vanish at $T_{\rm c}$, a phenomenon known as the normal state gap or pseudogap\cite{Timusk1999}. This characteristic led Emery and Kivelson\cite{Emery1995} to propose that underdoped cuprates are limited by small superconducting phase stiffness. Consequently, this results in a regime with a superconducting gap but significant phase fluctuations\cite{tconset7}, which helps to explain the presence of the pseudogap above $T_{\rm c}$.

Similarly, the Hall coefficient $R_{\rm H} = 1/n e$, where $e$ is the electron charge and $n$ is the itinerant carrier density, is negative and constant in metals due to the conduction band density. In contrast, in high-temperature superconductors (HTS), it is positive because of the presence of planar holes and exhibits a puzzling significant variation with temperature \cite{Hall.1994, Hall.2004, Hall.2006, Hall.2007}. Previous measurements of $R_{\rm H}$ on La$_{2-x}$Sr$_x$CuO$_4$ reported in Ref.\citenum{Hall.2007} were fitted using thermally activated excitation gaps across the relevant bands and impurity states, often with unrealistically adjustable local hole densities, in some cases exceeding four holes per copper atom\cite{Hall.2007}. Gor'kov and Teitel'baum proposed a more realistic model\cite{Gorkov2006} that considered the sum of two components: a temperature-independent term $n_0(p)\propto x$ due to external strontium doping. However, to fit $R_{\rm H}(p,T)$, they made this term significantly greater than $x$ for $p\ge 0.1$, and added an exponential thermally activated contribution.

Before presenting our model for the $R_{\rm H}(p,T)$ we need to address the role of planar charge instabilities \cite{Kivelson1998}, which have been observed in various forms, such as stripe phases\cite{Tranquada1995a}, checkerboard order\cite{Hanaguri2004}, puddles\cite{Lang2002}, and nematic phases \cite{Wu2017}, across several materials and compounds \cite{Comin2016}. Initially, this phenomenon was attributed solely to weakly doped compounds, particularly around $p = 1/8$ Cu atom, where the CDW, or incommensurate charge order (CO) is more intense\cite{Ghiringhelli2012,Huecker2014,Blanco-Canosa2014,Loret2019}, while overdoped compounds were traditionally considered to be Fermi liquid materials. However, this perspective is evolving due to numerous recent discoveries of CDW phases and local inhomogeneities in overdoped La$_{2-x}$Sr$_x$CuO$_4$ (LSCO), with doping levels reaching $p \equiv x = 0.21$\cite{Wu2017,Fei2019,Tranquada2021} and potentially up to $p = 0.27$ in the form of puddles\cite{OverJJ2022}. Furthermore, the observed onset of diamagnetism\cite{Over2023} above the $T_{\rm c}$ for LSCO with $p \ge 0.25$ exhibits a striking similarity to fluctuating superconductivity around optimal doping, which occurs below the onset temperature $T_{\rm c}^{\rm max}$, observed in the Nernst effect\cite{Wang2006,Nernst2010,tconset1,tconset2,tconset3}, in tunnelling\cite{Tunnel2000,Yurgens2003} and angle-resolved photoemission spectroscopy (ARPES)\cite{Kanigel2008,Vishik2014,NodalGap2015,Shen2019}, has garnered significant attention. More recently, charge density waves (CDW) have also been detected at the opposite end of the phase diagram, in the low-doping insulator region\cite{Insul.Stripes2019}. Even more surprisingly, CDW has been observed in the undoped Mott insulator parent compound\cite{Kang2023}.

Given all these experimental observations, a general theory of cuprates must account these ubiquitous charge instabilities\cite{tps5}, the strong phase fluctuations above $T_{\rm c}$, the pseudogap, and specific properties such as the Hall effect, which is discussed in detail here. We recall that early $R_{\rm H}$ measurements on  La$_{2-x-y}$Nd$_y$Sr$_x$CuO$_4$
attributed the rapid decrease in the magnitude of the Hall coefficient at low temperatures
to an evidence for one-dimensional spin-charge ordered stripe phase\cite{Noda1999}. This interpretation
suggests that a mechanism accounting to CDW/CO
formation is in the heart of the HTS properties.
Therefore, we propose that this general theory may be based on mesoscopic electronic phase separation, simulated by the theory of phase-ordering dynamics. Specifically, we employ the time-dependent non-linear Cahn-Hilliard (CH) differential equation, which is typically used to describe the continuous order-disorder transition of a binary alloy\cite{Cahn1958}.

It is important to emphasize that the Cahn-Hilliard (CH) approach characterizes the dynamical phase separation (PS)
process through the phenomenological segregation of local hole density, without distinguishing whether the CDW involve the carriers
in the O$_{2p}$ orbitals, as experimentally observed\cite{Comin2016,Fujita2014,Achkar2016}, or in the Cu$_{3d}$ orbitals.
This is because we assume that each unit cell may contain a local hole density.
	
We show that the calculations presented here provide an almost perfect reproduction of the $R_{\rm H}(p, T)$
using the values of the pseudogap or $T^*(p)$, the maximum superconducting fluctuations energy $T_{\rm c}^{\rm max}(p)$,
and the Ginzburg-Landau (GL) phase separation free energy amplitude that is also function of $p$.

\section{The Cahn-Hilliard differential equation}

The CH equation may be derived by minimizing a Ginzburg-Landau (GL) free energy 
expressed through a power expansion of the phase separation order parameter \cite{Bray1994}. 
The calculations start with the time-dependent CH equation, which is formulated in terms of 
the local phase separation order parameter $u (r_i,t) = (p(r_i,t) - p)/p$, 
where $p( r_i,t)$ represents the local charge or hole density at a unit cell position $i$ in the CuO 
plane, and $p$ denotes the average doping level of strontium (Sr). The variable and ``$t$'' 
indicates the simulation time\cite{Mello2017,Mello2021,Mello2020a,Mello2020c,Mello2022}. 
The simulation evolves to total phase separation, but it is 
stopped when the charges reach a configuration close to a given one, i.e., with 
the observed\cite{Comin2016} $\lambda_{\rm CO}$ of a sample. The CH equation 
is derived by minimizing the GL free energy\cite{deMelloKasal2012,deMello2014,Mello2017}: 
\begin{equation}
f(u)= {{\frac{1}{2}\varepsilon |\nabla u|^2 + V_{\rm GL}(u,T)}},
\label{FE}
\end{equation}
where $\varepsilon$ is the parameter that controls the charge modulations $\lambda_{\rm CO}$.The expression ${V_{\rm GL}}(u,T)= -A^2u^2/2+B^2u^4/4+...]$ represents a double-well potential. Charge oscillations occur below the electronic phase separation temperature, denoted as $T_{\rm PS}$, where $B = 1$, and $A^2 = \alpha(p) [T_{\rm PS}-T]$.

\begin{figure}[!ht] 
\centerline{\includegraphics[width=87mm]{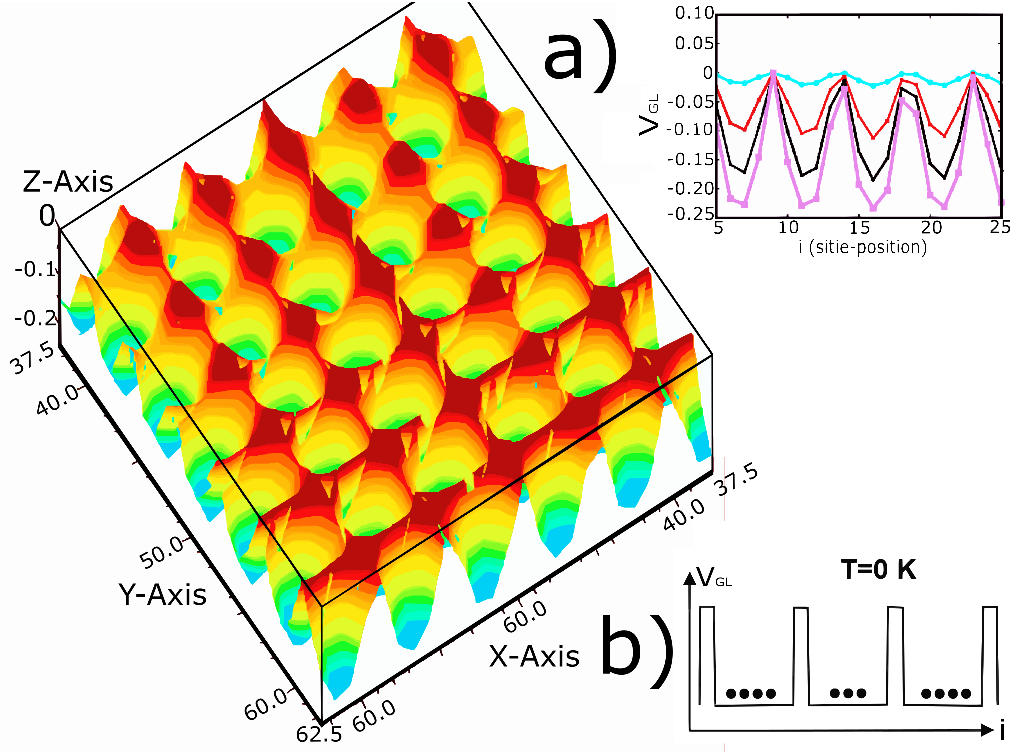}}
\caption{ Low-temperature CH simulations of the GL potential, which consists of an array of wells that gives rise to
the CDW phase of the $p = 0.14$ LSCO\cite{Mello2020a,Mello2021}.  The figure is from the central part of a
CuO planar $100 \times 100$ sites free energy simulation. Inset (a) displays the $V_{\rm GL}$ profile along the $x$-direction
for increasing temperatures, ranging from $T = 0$ K with maximum amplitude up to near $T^* (p) = T_{\rm PS} (p)$ with
vanishing amplitude, at which point the system becomes
uniform and the GL potential modulations disappear.  The low-temperature value is utilized in the quantum mechanical transmission
coefficient probability\cite{QMGasiorowicz} (Eq. \ref{eq4}) for calculating the temperature dependence of the Hall effect.
In inset (b), we present the simplified potential barriers used for the $R_{\rm H}$ calculations at $T = 0$ K.}
\label{fig1}
\end{figure}

We solved the nonlinear CH differential equation in the CuO plane using a stable and fast conservative finite difference scheme\cite{Otton2005}. In Fig.\ref{fig1}, we present a typical low-temperature simulation of $V_{\rm GL} (u(r_i, T))$ and its wells, which tend to confine carriers in alternating high- and low-density domains, giving rise to the CDW with wavelength $\lambda_{\rm CO}$. In the inset (a), the amplitude of $V_{\rm GL} (u(r_i, T))$ along the $x$-direction diminishes as the temperature approaches $T_{\rm PS} \approx T^*$, beyond which the system becomes homogeneous. 

The average height of the barriers between the two types of wells that host the two phases (high and low charge density) is a significant quantity and is proportional\cite{Mello2020a,Mello2020c,Mello2022} to $A^4/B^2$. Specifically, we have $\langle V_{\rm GL}(p, T) \rangle = \alpha(p)^2 [T^* - T]^2$ for $T \le T^*$. The quantity $\langle V_{\rm GL}(p, T) \rangle$ determines the charge modulations and, consequently, the mobility of the carriers. 

We illustrate a typical result for the case of $p = 0.14$ in LSCO in Fig. \ref{fig1}. As $T$ approaches $T^*(p) = T_{\rm PS}$, the amplitude of $V_{\rm GL}$ or the CO modulation decreases, as shown in the inset (a) of Fig. \ref{fig1}. This decrease leads, for instance, to the softening of the x-ray peaks\cite{Huecker2014,Blanco-Canosa2014} and the tunnelling signal\cite{Tunnel2000,Yurgens2003}.

On the other hand, charge localization into CDW patterns reduces the average kinetic energy and induces ion-mediated superconducting pair interactions \cite{deMelloKasal2012,Mello2020a,Mello2020c}. Such indirect interactions may yield local superconducting pair amplitudes, which we calculated using a self-consistent Bogoliubov-deGennes (BdG) approach with an attractive potential proportional to $\langle V_{\rm GL}(p, T) \rangle$, along with a method that preserves the CDW structure constant\cite{DeMello2012}. This aligns with the established understanding that the typical HTS coherence length is smaller than the CO wavelength, i.e., $\xi_{\rm{sc}} \le \lambda_{\rm{CO}}$. In this framework, akin to the Emery and Kivelson theory\cite{Emery1995}, superconducting long-range order is established through Josephson coupling\cite{deMello2014,Mello2017,Mello2020a,Mello2021,Mello2022} between the order parameters in the charge domains. 

 When the temperature is risen, the $V_{\rm GL}(p,T)$ modulations decrease, as illustrated in inset (a) of Fig. \ref{fig1},
the charge domains melt at $T^* (p)$ and the carrier density becomes uniform, resulting in the $R_H(p,T)$ high-temperature
saturation. Since $T^* (p)$ is very large in the low doping regime, this saturation is easily seen in the overdoped compounds
(see Fig. 4 of Ref. {\citenum{Hall.2007}} and our Fig. \ref{fig2}).

\section{The model to the Hall coefficient}

We now apply these ideas to interpret the Hall coefficient $R_H(p, T)$ measurements \cite{Hall.1994,Hall.2004,Hall.2006,Hall.2007},
beginning with the Mott insulator, $p = 0$ or $n = 1$, and the insulating region with $p \leq 0.05$, extending to overdoped
compounds with $p > 0.16$ using a unified model. The primary idea is that the CO or CDW modulations
are responsible for the significant temperature dependence of $R_H(p, T)$. As the temperature increases, the CDW modulations flatten,
as depicted in Fig. \ref{fig1}(a), which free some carrier, increases the mobility and consequently decreases 
$R_H(p ,T)$, a generally observed effect for $p \ge 0$.

In the Mott insulator region $R_H(p, T)$ changes smoothly with $p$ as it approaches the undoped (half-filled) compound. 
This is clearly seen in Fig. 1 of Ref. {\citenum{Hall.2007}} where the Hall coefficient
of $p = $ 0.0, 0.01, 0.02 and larger are plotted side-by-side suggesting a continuous behavior
approaching $p = 0.0$, which lead us to believe that all these compounds share similar properties.
Therefore, we assume that all the insulator region is dominated by the same phase separation (PS) mechanism
including charge and spin order at very low doping\cite{Insul.Stripes2019,Kang2023}, leading to
the possible existence of the CDW at half-filling.

Then, in a typical compound with $n = 1 + p$ carriers, the $n = 1$ holes self-organize into nearly static
modulations that shape the CDW/CO background because $T^*(p \approx 0)$ is very large. When strontium (Sr) doping increases, the additional holes are
guided by the $n = 1$ modulation and either agglomerate or segregate into alternating regions of high and
low densities, forming charge domains.

The resulting non-uniform density traps the more mobile $p$ charges, favoring the local Cooper pairs or superconducting amplitude,
increasing $R_H(p, T)$ at very low temperatures. As the temperature approaches $T_{\rm c}$, the Cooper pairs
begin to break down, leading to an increase in the density of carriers and a subsequent decrease in $R_H(p, T)$
after reaching a maximum.

In summary, $n = 1$ is the source of the CDW anomaly, while $p$ fills the low and high hole density domains and
are the source of Cooper pairs below the onset of pair formation at $T_{\rm c}^{\rm max}$. These two
types of carriers ($n = 1$ and $p$), which exhibit different dynamics, provide an interpretation for the two-gap or
two-particle scenario that has been identified by several different experiments\cite{Huefner2008, Wise2008, Yoshida2012, NodalGap2015}.

\subsection{Thermally Excited Hole Carriers}

Following the above discussion, we identify three distinct independent carrier contributions to
$R_{\rm H}(p,T)$: the $n = 1$ carriers that are essentially static at low temperatures and thermally excited with an energy scale proportional to $T^*(p)$,
the $p$ holes that tend to be condensed in the superconducting phase and are also thermally excited at the onset of pair formation temperature $T_{\rm c}^{\rm max}$,
and the carrier tunnelling through the free energy potential $V_{\rm GL} (T)$ barriers, shown in Fig. \ref{fig1} and in its insets.

It is important to note that we attribute the variation of $ R_{\rm H}(p = 0, T)$ with temperature \cite{Hall.2007} to the melting of the
CDW modulations present at half-filling ($n = 1$ and $p = 0$), and similarly in the insulating region, as recently confirmed
experimentally \cite{Insul.Stripes2019,Kang2023}. At high temperatures, the modulations of $V_{\rm GL}$
disappear near $T_{\rm PS}(p) \propto T^*(p)$, which is shown in the inset of Fig. \ref{fig1}(a). Accordingly, charge carriers,
proportional to $exp(-T^*(p)/T))$ are thermally excited, resulting in the following contribution to the hole density,
\begin{equation}
n_1(p,T) = n B_{\rm c} (1 + exp(-T^*(p)/T))
\label{eq2}
\end{equation}
where the half-filled $n = 1$ hole per Cu site is present for all compounds, and where $B_{\rm c} = 2$
is the number of CuO planes in the unit cell
of LSCO\cite{Shibauchi94}. We emphasize\cite{tps16,tps13,tps6} that Eq. \ref{eq2} is the same for all samples,
with their respective $T^*(p)$ listed in Table 1,
and compiled from many results (See Supplemental Material at [URL will be inserted by publisher] which compiles several experimental data points related to pseudogap temperatures ($T^{*}$) and the onset of pair formation ($T_{\rm c}^{\rm max}$), both of which are crucial to the model. See also references [\citenum{tps16,tps13,tps6,tps12,tps11,tconset5,tconset6,tconset8,tconset9,tps4,tps14,tps5}] therein).

\begin{table}[!ht]
\caption{ Properties of LSCO used in the calculations. The first column is 
hole doping density per unit cell. 
The second is $T^*$ and third is $T_{\rm c}$ which does not enter in the
calculations but is a good reference, and the last column is the renormalized free
energy barrier $\sqrt( \langle V_{\rm GL}\rangle')$ or roughness factor explained
in the text.
}
\begin{tabular}{|c|c|c|c|c|}\hline \hline
Sample &  $T^*$ (K) & $T_{\rm c}$ (K) & $T_{\rm c}^{\rm max}$ (K) 
& $\sqrt(\langle V_{\rm GL} \rangle')$   \\ \hline
 {\it p} = 0.00 & {\bf 1800} &  0.0 & {\bf 0.0 }& {\bf 9.0 }   \\ \hline
 {\it p} = 0.01 & {\bf 1700} & 0.0 & {\bf 10.0  }& {\bf 6.0} \\ \hline
 {\it p} = 0.05 & {\bf 1350} & 1.0  &{\bf 70.0 }& {\bf 4.0} \\ \hline
 {\it p} = 0.08 & {\bf 1200}  & 20.0 &{\bf 225.0} &{\bf 2.95} \\ \hline
 {\it p} = 0.12 & {\bf 900} & 34.0 &{\bf 190.0 }&  {\bf 2.30} \\ \hline
 {\it p} = 0.15 & {\bf 850} & 39.0  &{\bf 150.0  }& {\bf 1.8}\\ \hline
 {\it p} = 0.18 & {\bf 600} & 36.0  & {\bf 100.0 } & {\bf 1.1}\\ \hline
 {\it p} = 0.21 & {\bf 300} & 24.0  &{\bf  80.0  }& {\bf 0.6}\\ \hline
 {\it p} = 0.23 & {\bf 140} & 18.0  &{\bf  70.0 } & {\bf 0.4}\\ \hline
 {\it p} = 0.25 & {\bf 80} & 12.0  & {\bf 40.0 } & {\bf 0.2}\\ \hline
 \end{tabular}
 \label{table1}
 \end{table}

Secondly, the $p$ holes of a compound with $n = 1 + p$ form the CDW domains and local superconducting pairs within the
charge domains\cite{Mello2020a,Mello2021,Mello2022,Mello2023}.
The energy of these pairs is not scaled by $T_{\rm c}(p)$ but rather by 
$T_{\rm c}^{\rm max}(p)$. This refers to
the onset of superconducting pairs or precursor pairs, which have been detected by numerous experiments, such as
tunnelling\cite{Tunnel2000,Yurgens2003}, Nernst effect\cite{Wang2006}, ARPES\cite{Kanigel2008} and STM \cite{Gomes2007},
along with other references cited in the supplementary information (SI). At low temperatures
within the superconducting regime, the Cooper pairs do not contribute to $R_{\rm H}$. When the temperature increases,
these pairs eventually break apart, resulting in an increase in the number of normal carriers, and $R_{\rm H}$ reaches a
maximum between $T_{\rm c}(p)$ and $T_{\rm c}^{\rm max}(p)$. Consequently, the number of unbound holes generated
from the breaking of superconducting local pairs are also influenced by thermal excitation, scaled by $T_{\rm c}^{\rm max}$;
\begin{equation}
n_2(p, T) = p B_{\rm c} (1 + exp(-T_{\rm c}^{\rm max}(p)/T)). 
\label{eq3}
\end{equation}

The values of $T_{\rm c}^{\rm max}(p) $ used in the calculations are listed in Table I. They are averages derived from a compilation
of numerous measurements, which are detailed in the SI. These thermally activated phenomenological expressions
are similar to the Gor`kov and Teitel`baum\cite{Gorkov2006} proposal.

\begin{figure}[!h]
\centerline{\includegraphics[width=88mm]{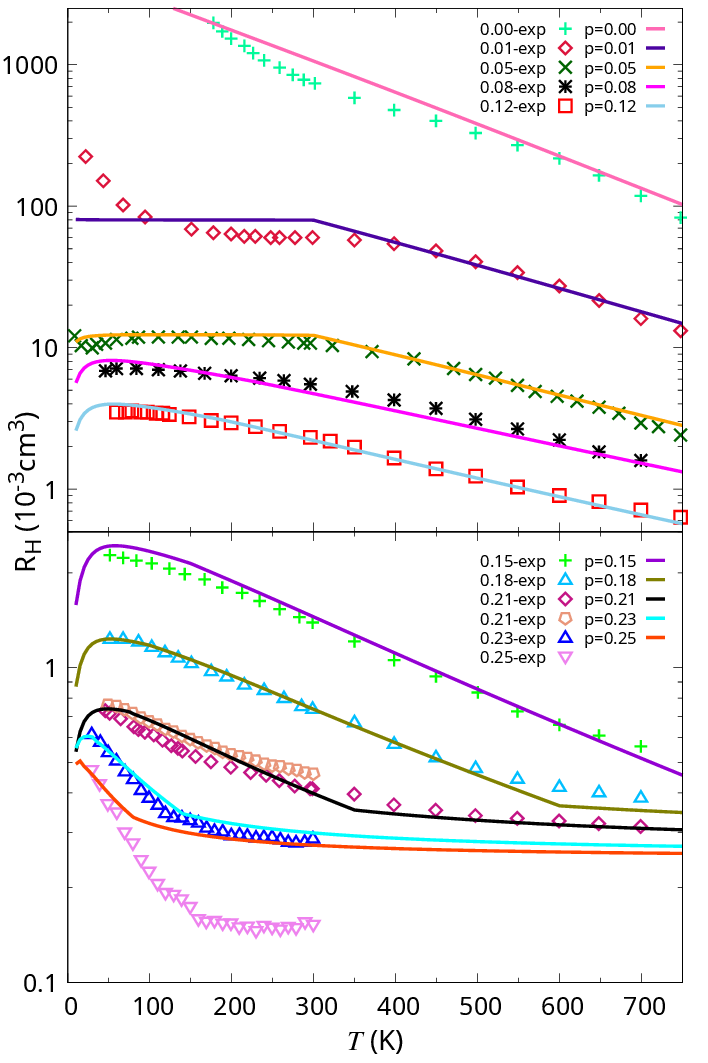}}
\caption{
The experimental Hall effect and theory are presented together. Data points from Ref. {\citenum{Hall.2007}} and $p=0.21$ from Ref. {\citenum{Hall.2004}}
are compared with the theoretical model (Eq. \ref{eq5}) derived in the text, represented by continuous lines. The agreement is satisfactory for values of
$p$ ranging from $p = 0.0$ to 0.23. However, the case of $p = 0.25$ is the only instance that yields significantly larger values above $T = 100$ K.
This discrepancy is likely attributed to the negative $R_{\rm H}(p, T)$ of out-of-plane electrons in strongly overdoped samples, as noted in
Ref. {\citenum{Hall.2006}}.}
\label{fig2}
\end{figure}

The final and most original effect on the $R_{\rm H}(p, T)$ is the background modulation of \( V_{\rm GL}(p, T) \) \cite{Mello2020a},
which creates barriers and valleys that affect carrier mobility. As shown in the inset (a) of Fig. \ref{fig1},
the amplitude of this modulation is larger at low temperatures but decreases as the temperature rises. Its precise
temperature dependence is discussed following the GL free energy (Eq. \ref{FE}). Consequently, the motion of free carriers
under this modulated background is described by the tunnelling current probability between charge domains or grain boundaries.
This effect is challenging to calculate accurately, so we adopted a one-dimensional simplified model of the quantum mechanical transmission coefficient\cite{QMGasiorowicz}. In this context, the tunnelling probabilities through rectangular barriers
are illustrated in the inset (b) of Fig. \ref{fig1}, with a width of $\lambda_{\rm CO}/2$ and a height proportional to $\langle V_{\rm GL}(p, T)\rangle{(1 - (T/T^*))^2}^{1/2}$, is given by\cite{QMGasiorowicz}:
\begin{equation}
T_{\rm trans}^2 = e^{ -[{\gamma\langle V_{\rm GL}\rangle (1 - (T/T^*))^2}]^{1/2}}
\label{eq4}
\end{equation}
where $\gamma = 2m\lambda^2_{\rm CO}/\hbar^2$ and $m$ is the electron mass. 

The values of $ \langle V_{\rm GL}\rangle$ have been previously calculated\cite{Mello2021,Mello2022} and
are presented in Table \ref{table1}. It is important to note that $\gamma \langle V_{\rm GL}\rangle$ is a
dimensionless measure of the CDW roughness and exhibits a doping dependence similar to that of the pseudogap,
according Fig. 2 of Ref. \citenum{Mello2020a}.

Now, the total hole density measured in the Hall experiments are
\begin{equation}
n (p, T)= (n_1(p, T)+n_2(p, T)) T_{\rm trans}^2 
\label{eq5}
\end{equation}
that contains all the temperature dependence of the number of holes, which has been the main 
and most puzzling feature of $R_{\rm H}(p, T) = 1/(e n(p,T))$. 

We emphasize that these simple expressions (Eqs. 2-5) are robust enough to describe 
{\it all} $ 0 \le p \le 0.25$ experimental values\cite{Hall.1994,Hall.2004,Hall.2006,Hall.2007} using
only their $T^*(p)$, $T_{\rm c}^{\rm max}(p)$ and the transmission coefficient $T_{\rm trans}^2$.
For $0.0 \le p \le 0.05$,
we keep the $n (p,T)$ constant for $T \le T_{\rm AF} \approx 300$ K, the N\'eel temperature, 
because of the antiferromagnetic (AF) insulator phase reduces drastically the carrier's mobility,
which yields a plateau in the $R_{\rm H}(p, T)$ data\cite{Hall.2007} .

The values of  $\langle V_{\rm GL}\rangle '$ $= 2m\lambda^2_{\rm CO}/\hbar^2\langle V_{\rm GL}\rangle$ = $\gamma \langle V_{\rm GL}\rangle$
in Eq. \ref{eq4} above, for under and overdoped LSCO are obtained from the tables in the Supplementary Information of
Ref.  {\citenum{Mello2021}} and $\lambda^2_{\rm CO}(p)$ from
Ref.  {\citenum{Comin2016} }. For instance, the optimally doped $p = 0.16$ has\cite{Mello2020a,Mello2021} $\langle V_{\rm GL}\rangle = 0. 234$
eV and $\lambda_{\rm CO} = 3.9 a_0$ with $a_0 = 3.78 \AA $  and $\sqrt{\gamma \langle V_{\rm GL}\rangle} = 1.833$, and
therefore we use in Eq. \ref{eq4}, the value 1.80 to $p = 0.15$. The $\sqrt \langle V_{Gl} \rangle^{'}$ to other
compounds are listed in the last column of Table \ref{table1}.

The PS GL free energy simulations (Fig.\ref{fig1}) show how a spatial uniform charge distribution changes or are shaped into stripes,
or any form of CDW at low temperatures. As mentioned, such an inhomogeneous charge background favors superconducting amplitude fluctuations\cite{DeMello2012,deMello2014,Mello2020a} below $T_{\rm c}^{\rm max}(p)$. An important consequence is that, for
low-doping compounds, physical properties such as superfluid density or quantum oscillations\cite{Hall2016,Mello2020c}
are dominated almost exclusively by the thermally excitation $p$ holes because $T_{\rm c}^{\rm max}(p) \ll T^*(p)$.

This feature implies that the mobile $p$ holes are the main protagonist of the low doping and low temperature physical properties.
However, for $p \approx 0.19$ or higher, $T_{\rm c}^{\rm max}(p)$ and $T^*(p)$ are of the same order of magnitude
(see Table \ref{table1}), and these carriers become indistinguishable; specifically, the semi static $n = 1$
carriers and the more mobile $p$ holes mix together. Consequently, the measured concentration in some
experiments\cite{Tallon2001,Cooper2009} crosses over from $p$ at low doping to $1 + p$ at high doping.
More recently, measurements of anisotropic
resistivity in LSCO observed\cite{Tranquada24} the strange-metal behavior for crystals with doped-hole
concentration $p = x > 0.19$, in contrast with the nonmetallic behavior for $p < 0.19$.
Similarly, detailed ARPES studies\cite{Shen2019} documented a sudden transition from a
strange metal, characterized by an incoherent spectral function, to a more conventional metal with well-defined quasiparticles
near $p \approx 0.19$.
In other words, properties like the CDW and superfluid density are contributed by the $p$ holes in the background of
$n = 1$ modulations, which soften near this $p \approx 0.19$ critical doping. Another consequence of this CDW softening is that, for a
long time CDW/CO order was believed to exist solely in the underdoped regime and around $p = 1/8$.

This intriguing result was also directly observed in Hall effect experiments at high magnetic fields with Y123
compounds\cite{Hall2016} and was attributed to Fermi surface reconstruction with
the development of electron pockets below $p \approx 0.16$. The same argument was used to explain the
drop of $R_H(p, T)$ into negative values in very high fields\cite{LeBoeuf2007}. However, ARPES experiments\cite{FermiArc2008} with
Y123 found only evidence of Fermi arcs,
suggesting that high magnetic fields might be inducing a state different from that being
studied in zero field. We recall that the Fermi arcs measured by ARPES\cite{FermiArc2008} can be explained as a $k$-space manifestation
of the real space phase separation produced by the CH phase separation\cite{Mello2020c}.

Therefore, taking all these factors into consideration, the Hall coefficients $R_{\rm H}(p,T)$ from Eq. \ref{eq5} are calculated
using a simple Fortran program with the experimental values of $T^*(p)$, $T_{\rm c}^{\rm max}$ and
estimated $\langle V_{\rm GL}\rangle $ listed in Table I. The results are presented in Fig. \ref{fig2}, alongside the
experimental data from Ref. \citenum{Hall.2007}, demonstrating a strong agreement from \( p = 0.0 \) to \( 0.23 \).
For $p \le 0.05$ there is a divergence at very low temperatures because the
charge variation with the temperature is essentially zero causing $R_H(p, T)$ to diverge, and
this behavior at $T \rightarrow 0$ is clearly not capture by our $T > 0$ K thermal excitation model
that is more accurate at finite temperatures.
On the other side, for the overdoped case at $p = 0.25$, the results diverge from our calculations above $T = 100$ K.
This discrepancy is most likely attributed to the negative Hall coefficient of out-of-plane electrons in strongly overdoped samples, as reported by Ref. \citenum{Hall.2006}.

\section{Conclusion}

 We emphasize that our single model accurately reproduces all the $R_{\rm H}(p, T)$ data, ranging from undoped Mott insulator to overdoped samples, and from low to high temperatures, using a straightforward expression (Eq. \ref{eq5}). The key point is that the increase in nominal charge density with temperature is primarily a consequence of the melting of CDW/CO charge modulations. Consequently, we identify the pseudogap temperature $T^*(p)$ as
the onset of charge instabilities or phase separation like the most 
significant energy scale in cuprates. Previously, we noted that $\langle V_{\rm{GL}} \rangle$ exhibits a doping dependence\cite{tps15} similar to that of $T^*(p)$ (as shown in Fig. 2 of Ref. \citenum{Mello2020a}). This finding is fully consistent with Raman spectroscopy studies\cite{Loret2019}, which confirmed the correlation between the doping dependence of $T^*(p)$, the CDW energy, and the preformed superconducting gap.

Another original finding from our calculations of $R_{\rm H}(p,T)$ is the presence of the CDW at half-filling ($n = 1$), which appears to intrinsically influence the charge instabilities that persist up to overdoped regime. Additional $p$ holes introduced by strontium (Sr) doping populate
alternate charge domains, resulting in the detection of two types of carriers previously identified\cite{Huefner2008}.
First, there are the less mobile carriers at half-filling ($n = 1$), which shape the CDW in all compounds.
Second, there are the mobile $p$ holes that are distributed across alternating CO domains. These two types of carriers were
distinguished by their different energies and properties, which converge near the crossover at $p \approx 0.19$. This phenomenon
has previously been attributed to a quantum critical point by different techniques\cite{Tallon2001,Cooper2009,Keimer2015,Hall2016,Shen2019,Tranquada24} and 
our present work offers a clear interpretation in terms of the two distinct energy scales. 
The new results discussed above are general
and provide an original
phenomenological approach to interpreting several non conventional HTS properties.

\section{Acknowledgement}

We acknowledge partial support from the Brazilian agencies CNPq and by Funda\c{c}\~ao Carlos Chagas Filho de Amparo Pesquisa do Estado do Rio de Janeiro (FAPERJ), Project No. E-26/211.270/2021.

%
\end{document}